%% file: main.tex
\documentclass{article}

\input{preamble/preamble}

\usepackage{spconf}
\input{preamble/preamble_acronyms}

\input{preamble/preamble_math}

\input{preamble/preamble_tikz}

\title{Direct optimization of F-measure for retrieval-based personal question answering}
\name{
\small{Rasool Fakoor$^\dagger$\thanks{ $\dagger$ Corresponding author },
Amanjit Kainth$^*$,
Siamak Shakeri$^*$,
Christopher Winestock,
Abdel-rahman Mohamed,
Ruhi Sarikaya\thanks{* Equal contribution as a second author}}}
\address{Amazon}
\begin{document}
%\ninept
%
\maketitle
\input{abstract}
\input{intro}
%\input{relatedworks}
\input{problem}
\input{contribution}

%\input{unsupervised}
\input{experiments}
\input{conclusion}

% all our results and tables stored in `tables.tex` below,
% don't include it in our final paper version.
% Only for OUR use.
%\input{tables}

% \input{figure}

\vfill
\pagebreak
\clearpage

% \input{cprt_forms}
% \input{references}

\nocite{*}
\bibliographystyle{IEEEbib}
\bibliography{refs}

\end{document}

%% file: preamble/preamble.tex
\usepackage[T1]{fontenc}
\usepackage{tgtermes}

\usepackage{amssymb}
\newcommand{\mathbold}[1]{\ensuremath{\boldsymbol{\mathbf{#1}}}}

\usepackage{times}

\usepackage{amsmath, amsthm, amsfonts}
\usepackage{mathtools}

\usepackage[
  paper  = letterpaper,
  left   = 1.25in,
  right  = 1.25in,
  top    = 1.0in,
  bottom = 1.0in,
  ]{geometry}
\usepackage[english]{babel}
\usepackage{afterpage}
\usepackage{enumitem}
\usepackage{framed}
\usepackage{xspace}

\usepackage{lineno}

\usepackage{ragged2e}

\newcounter{parcount}

\usepackage[usenames,dvipsnames]{xcolor}
\definecolor{shadecolor}{gray}{0.9}

\usepackage{graphicx}
\usepackage[labelfont=bf]{caption}
\usepackage[format=hang]{subcaption}

\usepackage{booktabs, array}
\usepackage{multirow}
\usepackage{diagbox}

\usepackage[algoruled]{algorithm2e}
\usepackage{listings}
\usepackage{fancyvrb}
\fvset{fontsize=\small}

\usepackage[colorlinks,linktoc=all]{hyperref}
\usepackage[all]{hypcap}
\hypersetup{citecolor=MidnightBlue}
\hypersetup{linkcolor=MidnightBlue}
\hypersetup{urlcolor=MidnightBlue}
\usepackage[nameinlink]{cleveref}
\creflabelformat{equation}{#2\textup{#1}#3}  % <- remove parenthesis from equations

\usepackage
[acronym,smallcaps,nowarn,section,nonumberlist]{glossaries}
\glsdisablehyper{}

\definecolor{strings}{rgb}{.624,.251,.259}
\definecolor{keywords}{rgb}{.224,.451,.686}
\definecolor{comment}{rgb}{.322,.451,.322}

\lstdefinelanguage{python}{
  keywords=[3]{Normal, Bernoulli, Beta, Categorical, Dirichlet,
  Exponential, MultivariateNormalFull, RandomVariable,
  DirichletProcess, Empirical, PointMass, Gamma,
  MAP, Inference, KLqp, HMC, SGLD, KLpq,
  VariationalInference, MonteCarlo, ConjugateInference, GANInference,
  rnn_cell, dirichlet_process, cond, body, generative_network,
  discriminative_network,
  evaluate, ppc, copy, dot, get_session},
  morecomment=[l]{\#},
  morecomment=[s]{"""}{"""},
  morestring=[b]',
  morestring=[b]",
  alsoletter={<>=-+/*},
  sensitive=true
}

\renewcommand{\texttt}[1]{\lstinline[basicstyle=\fontsize{8pt}{8.25pt}\selectfont\ttfamily]{#1}}

%% file: preamble/preamble_acronyms.tex
\newacronym{PPL}{ppl}{probabilistic programming language}
\newacronym{GPU}{\textnormal{\uppercase{gpu}}}{graphics processing unit}

\newacronym{VAE}{vae}{variational auto-encoder}
\newacronym{RBM}{rbm}{restricted Boltzmann machine}
\newacronym{DLGM}{dlgm}{deep latent Gaussian model}
\newacronym{GAN}{gan}{generative adversarial network}
\newacronym{RNN}{rnn}{recurrent neural network}

\newacronym{VI}{vi}{variational inference}
\newacronym{MCMC}{mcmc}{Markov chain Monte Carlo}
\newacronym{MC}{mc}{Monte Carlo}
\newacronym{HMC}{hmc}{Hamiltonian Monte Carlo}
\newacronym{MAP}{map}{maximum a posteriori}
\newacronym{IWAE}{iwae}{importance-weighted auto-encoder}
\newacronym{HVM}{hvm}{hierarchical variational model}
\newacronym{KL}{kl}{Kullback-Leibler}
\newacronym{ELBO}{elbo}{\emph{evidence lower bound}}

\newacronym{BOW}{bow}{Bag-of-Words}
\newacronym{MLE}{mle}{maximum likelihood estimation}
\newacronym{QA}{qa}{question answering}

%% file: preamble/preamble_math.tex
\newcommand{\mbv}{\mathbold{v}}

%% file: preamble/preamble_tikz.tex
\usepackage{tikz}
\usetikzlibrary{bayesnet}

\pgfdeclarelayer{edgelayer}
\pgfdeclarelayer{nodelayer}
\pgfsetlayers{edgelayer,nodelayer,main}

\definecolor{hexcolor0xbfbfbf}{rgb}{0.749,0.749,0.749}

\tikzset{>=latex}
\tikzstyle{none}   = [inner sep=0pt]

\tikzstyle{line}  = [ - ]
\tikzstyle{arrow}  = [ ->, shorten <=1pt, shorten >=1pt ]
\tikzstyle{ardash} = [ dotted, ->, shorten <=1pt, shorten >=1pt ]

\tikzstyle{empty}=[circle,opacity=0.0,text opacity=1.0,inner sep=0pt,minimum
width=0pt,minimum height=0pt]
\tikzstyle{box}=[rectangle,fill=White,draw=Black]
\tikzstyle{filled}=[circle,fill=hexcolor0xbfbfbf,draw=Black]
\tikzstyle{hollow}=[circle,fill=White,draw=Black]
\tikzstyle{param}=[rectangle,fill=Black,draw=Black,inner sep=0pt,minimum width=4pt,minimum height=4pt]

%% file: abstract.tex
\begin{abstract}

Recent advances in spoken language technologies and the introduction of many customer facing products, have given rise to a wide customer reliance on smart personal assistants for many of their daily tasks. In this paper, we present a system to reduce users' cognitive load by extending personal assistants with long-term personal memory where users can store and retrieve by voice, arbitrary pieces of information. The problem is framed as a neural retrieval based question answering system where answers are selected from previously stored user memories. We propose to directly optimize the end-to-end retrieval performance, measured by the F1-score, using reinforcement learning, leading to better performance on our experimental test set(s). 
\end{abstract}
%

%For best end-to-end performance, we directly optimization the overall system retrieval F-measure performance using both, a smoothed variant and the reinforce algorithm leading to better performance across our experimental test sets. Many experiments and performance analysis are presented to study the overall impact of semi-supervised learning and the impact of sub-word input representation for mitigating ASR errors in the input memories and questions.

%% file: intro.tex
\section{Introduction}
\label{sec:intro}

Recent advances in speech recognition~\cite{hinton2012deep,DahlAcero}, speech enhancement~\cite{rf2017NIPSW, rfICASSP2018}, natural language understanding~\cite{zettlemoyer2012learning,Mori2008}, question answering~\cite{weston2015towards, Xiong2016DynamicCN, Sordoni2016IterativeAN}, and dialogue systems~\cite{young2010hidden, LiMRJGG16} have fueled the current surge in research and development for smart personal assistants~\cite{Sarikaya} like Alexa, Siri, Google assistant, and Cortana, with many use cases around shopping, music, etc.

In this paper we present a system for providing personal assistants a long term personal memory that enable users to store anything they want to remember by voice, and then later ask questions about it. An example use case is shown in Table \ref{table:ex_qa}. This system extends long-term memories of users and enables them to store and retrieve arbitrary pieces of information they are juggling in their minds.

\begin{table}[t]
\begin{tabular}{lc|}
 \hline
\textbf{Question}: \\
what did i do with ben's cell phone \\
\textbf{Answers}: \\
1. \textcolor{Green}{$\checkmark$} i gave benny's cell in for repairs at the \\
 store on first street\\
2. \textcolor{Green}{$\checkmark$} i left ben's iphone on the kitchen table\\
3. \textcolor{Green}{$\checkmark$} i sent bennie's old phone to mat \\
4. \textcolor{Red}{$\times$} ben wants a new cell phone for his birthday\\
5. \textcolor{Red}{$\times$} dad's cell is an iphone eight\\
6. \textcolor{Red}{$\times$} the screen of benjamin's phone is broken\\
 \hline
\end{tabular}
\caption{An of example QA group. These samples are output of an ASR system and hence normalized, i.e. noisy data, no punctuation, no capitalization, etc.
}\label{table:ex_qa}
\end{table}

The system is framed as a \acrlong{QA} (QA) system over user generated memories which is related to QA systems with answers extracted from unstructured public sources like Wikipedia \cite{rajpurkar2016squad}, Neural Information Retrieval \cite{mitra2017neural}, Text Matching \cite{wang2017bilateral}, and Machine Comprehension \cite{seo2016bidirectional}.

One of the challenges for retrieval models is that they are trained on criteria different from the needed business metrics, which are not usually differentiable, e.g. training on pairwise matching while the overall performance is measured by $F1$ score. For example, \cite{QuocOptim2007} proposed a method for direct optimization of the relevance loss functions in ranking problems via structured estimation in Hilbert spaces by formulating it as a linear assignment problem.

%Their model mainly focused on  directly minimization of the multivariate performance measures. Even though there are some interesting results in this paper, extending their method to deep model seems very challenging.

In \cite{PastorPellicer} the authors proposed a method to directly optimize a relaxed version of the F-measure which is similar to a variant we present in this paper (see \ref{subsec:smooth}). In \cite{Xu2016ExpectedFT}, the expected F-measure is used to train a neural parsing model on sentence-level $F1$.

%as a loss function.  This method is closely related to our smooth F1  approximation. However, our smooth F1 optimization has different structure as it measure the loss per one query and memory groups rather than an input pair. In addition, their goal was mainly to deal with imbalanced classes which is in our case it is one of the goals. Curriculum learning and mixture losses are other difference between our model and theirs.
Another way to deal with non-differentiable functions is to approximate the gradients using the REINFORCE algorithm \cite{Williams1992}. In \cite{RanzatoCAZ15, Rennie2017}, REINFORCE was used for sentence generation both in machine translation and video captioning tasks. The goal in both papers was to directly  optimize evaluation metrics of interest such as BLEU-4 or CIDEr.

In this paper, we focus on improving the overall performance by incorporating the $F1$ score as part of the optimization objective. The main contributions of the paper are:
 \begin{itemize}
 \item Introducing a system for spoken personal QA.
% \vspace*{-6pt}
 \item Proposing a method to directly optimize $F1$, and introducing stable optimization strategies.
 % \vspace*{-6pt}
 \item Present extensive empirical evidence and analysis discussing the viability of this approach, comparing it to traditional optimization techniques.
  \end{itemize}

  The paper is organized as follows:
  Section \ref{sec:problem} describes the problem and challenges associated with it. Sections \ref{sec:contrib} and \ref{sec:optimization} describe
  our proposed models and optimization schemes. Finally, Sections \ref{sec:experiments} and \ref{sec:analysis} discuss experimental results.
%Section \ref{sec:semisupervised} talks about utilizing semi-supervised data and fi

%% file: problem.tex
\section{Personal Question Answering}
\label{sec:problem}

%In this paper we explore the problem of document matching using deep learning in a retrieval-based question-answering (QA) system. This QA system is part of an experimental voice-controlled digital assistant in which a user can ask the assistant to store personal information, i.e. one or more memories, and then retrieve the relevant memories at a later time using a question.

The problem is framed as a retrieval-based QA system over stored memories. More formally, given a question $q$ and a set of user memories $M_q$, the system returns a subset of memories $R_q$ which are relevant and answers the spoken question. Throughout the paper, the set $\{q, M_q\}$ is referred to as a QA group. Table \ref{table:ex_qa} shows an example QA group of four memories marked as relevant or irrelevant to the input question.
%the user asks $q = $ ``what did i do with ben's cell phone''. From the set of memories, $M_q = \{m_1, m_2, m_3, m_4\}$, the system should return the memories marked with a \textcolor{Green}{$\checkmark$}. The irrelevant memories, marked with a \textcolor{Red}{$\times$}, should not be returned. \\

\iffalse
\fbox{\begin{minipage}{23em}
\textbf{Question}:
what did i do with ben's cell phone  \\

\textbf{Answers}:
\begin{enumerate}[itemsep=-0.5mm]
  \item \textcolor{Green}{$\checkmark$} i gave benny's cell in for repairs at the store on first street
  \item \textcolor{Green}{$\checkmark$} i left ben's iphone on the kitchen table
  \item \textcolor{Green}{$\checkmark$} i sent bennie's old phone to mat
  \item \textcolor{Red}{$\times$} ben wants a new cell phone for his birthday
  \item \textcolor{Red}{$\times$} dad's cell is an iphone eight
  \item \textcolor{Red}{$\times$} the screen of benjamin's phone is broken
\end{enumerate}
\end{minipage}} \\\\
\fi
For solving this problem, a classification approach could be adopted  for the given question and memory pairs, but this approach is not optimal due to the large class imbalance between relevant and irrelevant user memories for each question as shown in Table \ref{table:relevant_label_percentages}. A better end-to-end formulation should take into account all user stored utterances (memories) when making relevance decisions for each individual memory, i.e. directly optimizing the $F1$ score for each QA group. However, it is challenging to directly optimize for $F1$ measure due to its discrete and non-differentiable nature.

Another challenge rises from the spoken nature of the presented system, where both user memories and questions are transcribed by an ASR system. Due to varying acoustic conditions, names or locations could be recognized as ambiguous entities during storing a memory and recalling it, which are temporally distant. We found that this effect compounds the effect of ASR errors on the end-to-end retrieval performance.

%% file: contribution.tex
\section{Neural retrieval models}
\label{sec:contrib}
%In this section we present the architecture of our models and explain our proposed new optimization objective function.

%\subsection{Model Architecture}
We propose various optimization objectives and model architectures for determining the relevance of stored memories given a question. None of our proposed architectures contain any recurrent units because fast and efficient inference is key to a seamless user experience. We focus our efforts on optimization in this work, and demonstrate the effects of carefully constructed optimization objectives, to train a relatively simple model architecture to achieve high performance.

% Each of our proposed architectures has different strengths and complexities. They differ in their ability to handle class label imbalance, speech recognition errors and ground truth data sparseness and so yield different results on datasets which vary along these lines.

The input to a model is a question, $q$, and its corresponding set of memories, $M_q = \{m_1, ..., m_{|M_q|}\}$. Each question and stored memory undergoes a string preprocessing step to clean up the text and tokenize the utterance into words. The utterance is then encoded using word-level representations, for input to the model. We also experiment with using compositional word embeddings \cite{kim2016character,jozefowicz2016exploring} to distill task-specific subword knowledge into our model. Specifically, for a given query $q = (w_1, ..., w_{|q|})$, we define its feature vector to be the matrix of word embeddings:

\begin{equation*}
E_q = [\mbv_{w_1}, ..., \mbv_{w_{|q|}}] \in \mathbb{R}^{|q| \times d}.
\end{equation*}
where $|q|$ denotes the length of the question in words and $d$ is the feature dimension. We use similar terminology for the vector representing a memory, which we write here as $E_{m_i} \in \mathbb{R}^{|m_i| \times d}$, where $|m_i|$ denotes the number of tokens in the memory. We also experiment with modifying the word embedding to include compositional word embeddings, as a way of leveraging the task-specific information present in our corpus. The compositional word embedding module generates word representations using character embeddings, following closely the architecture of the $\mathtt{CharCNN}$ model presented in \cite{jozefowicz2016exploring}. This module is jointly trained with the rest of the model using the task-specific objective. Each word can then be represented by the concatenation of its pre-trained word vector and the $\mathtt{CharCNN}$ embedding.

% For a given word $w = (c_1, ..., c_{|w|})$, let $\mbC_w \in \mathbb{R}^{|w| \times d}$ be the character-level representation of $w$, where $d$ is the dimension of the character embeddings, which is not necessarily the same as word vector dimensionality. We apply a convolution between $\mbC_w$ and a kernel $\mbK_i \in \mathbb{R}^{k \times d \times n}$ of width $k$ and $n$ output channels, followed by a non-linearity and max-pool layers:
% \vspace{-12pt}
% \begin{align*}
%   y_i &= \mbC_w \ast \mbK_i  + b_i \in \mathbb{R}^{\hat{|w|} \times n} \\
%   y_i &= \mathtt{maxpool}(\mathtt{relu}(y_i)) \in \mathbb{R}^n \\
%     y &= \mathtt{concat}(y_1, ..., y_f) \in \mathbb{R}^{f\cdot n} \\
%   \mby_w &= \mathtt{Linear}(y) \in \mathbb{R}^{d}
% \end{align*}
% We do this for $f$ kernels of varying width and concatenate the output of these $f$ convolutions before applying a final linear layer to produce embeddings of dimension $d$. We refer to this character-aware embedding model as $\mathtt{CharEmbedding}$ in our results.

In the next sections, we explain the architectures of the models $\mathtt{TEFF}$ and $\mathtt{TEFFCH}$\footnote{Model name(s) used for simplicity. }. % and subsequently the models $\mathtt{SARD}$ and $\mathtt{SARDCH}$, which employ attention.
 %The combined character-aware and pre-trained word embeddings is given by $\mbu_w = \mathtt{concat}(\mbv_w, \mby_w)$.
\subsection{$\mathtt{TEFF}$ and $\mathtt{TEFFCH}$}

The $\mathtt{TEFF}$ model is comprised of $N$ fully-connected layers, followed by a max-pool layer across time to produce a fixed-dimensional vector. The query and memory embeddings are both processed through the same network, producing $h$-dimensional vectors for the query and memory, represented by $u$ and $v$, respectively.
% \vspace*{-15pt}
% \begin{align*}
%   y_1 &= \mathtt{Linear_A}(E_q) \in \mathbb{R}^{|q| \times h} \\
%   y_2 &= \mathtt{dropout}(\mathtt{relu}(y_1)) \\
%   y_3 &= \mathtt{Linear_B}(y_2) \\
%   y_4 &= \mathtt{dropout}(\mathtt{relu}(y_3)) \\
% \end{align*}
The joint activation and logits are computed as:
\vspace*{-5pt}
\begin{align*}
  uv &= \mathtt{concat}(u, v, |u-v|, u\odot v) \\
  \mathtt{logits} &= \mathtt{softmax}(\mathtt{dropout}(uv)) \in \mathbb{R}^2 \\
\end{align*}
\vspace*{-30pt}

The $\mathtt{TEFFCH}$ model follows a similar structure, except the input embedding is given by the concatenation of the pre-trained word vector and the $\mathtt{CharCNN}$ embedding, that is jointly trained, producing an end-to-end model.

%\subsection{$\mathtt{SARD}$ and $\mathtt{SARDCH}$}

%Our $\mathtt{SARD}$ model is a neural attention model inspired by the Transformer component~\cite{Transfomer2017}. This architecture is composed of a stack of various neural layers: a highway layer ~\cite{Highway, SrivastavaGS15}, followed by a \textit{two way} multi-head self-attention layer (i.e. query $\Leftrightarrow$ memory), followed by position-wise feed-forward layers and layer normalization followed by fully-connected layers and then the output layer. Similar to $\mathtt{TEFFCH}$, the $\mathtt{SARDCH}$ model parametrizes the input embedding using the $\mathtt{CharCNN}$ module, while still retaining the same architecture as $\mathtt{SARD}$ thereafter.

\section{Direct optimization of F-measure}
\label{sec:optimization}

As the goal of our model is to correctly assign the label  `relevant' vs. `irrelevant' to a memory given a question, we can formulate the optimization objective as the maximization of labelling accuracy. Expressed as a loss function, we try to minimize the cross-entropy between the estimated class probabilities and the ground truth label distribution for a set of question-memory pairs:

\begin{equation}\label{eq:cs}
  \mathcal{L}_{cs}(\theta) = -\sum\limits_{i } \sum\limits_{j} ^{|M_q|} \sum_{c}  \text{log }p_{\theta}(m_j|q_i) \mathbb{I}_c (y_{ij})
\end{equation}
$m_j$ and $q_i$ denote the memory and question pair respectively, $y_{ij}$  is corresponding label,  $\mathbb{I}$ is the indicator function, and $p_{\theta}$ denotes the model, parametrized by $\theta$. Even though this formulation renders optimization straightforward, it leads to a discrepancy at evaluation time as we optimize for one objective during training but use another metric for model evaluation. More specifically, we optimize for maximum accuracy of question-memory-pair labelling during training but evaluate our model using the $F1$ score averaged across all QA groups. An analogous objective cannot be used as an optimization for $F1$, as $F1$ is not differentiable. Furthermore, Eq. \ref{eq:cs} does not address the large class imbalance between irrelevant memories versus relevant memories in a QA group.

To address this discrepancy, we propose a novel optimization objective that can directly estimate the evaluation metric.

\subsection{Policy Gradient based Approximation}\label{sec:pg}
Our goal is to maximize the expected $F1$ score, for a given dataset. To do this, we first formulate our task as a reinforcement learning problem in which our network  acts as the agent, i.e. policy network, and so provides the probability for taking a particular action on each question-memory pair:

\begin{equation}\label{eq:pg}
  \mathcal{L}_{\text{re}}(\theta) = -\mathbb{E}_{a^i \sim p_{\theta} (M_{q} | q)}[R(a^{M_q})]
\end{equation}
where $p_{\theta}$ is the policy network, $R(a^{M_q})$ is the reward function, and $a^i$ is the action given $(m_i, q)$. Since the reward is not differentiable, we use REINFORCE \cite{Williams1992, ZarembaS15} to estimate the gradients. Based on this algorithm, the gradients are calculated as follows:

\begin{equation}\label{eq:gpg}
  \nabla_{\theta} \mathcal{L}_{\text{re}}(\theta) = -\mathbb{E}_{a_i \sim p_{\theta}}[R(a^{M_q}) \cdot \nabla_{\theta} \text{log }p_{\theta} (M_q| q) ]
\end{equation}

Even though this new formulation has the potential to boost model performance, it presents several challenges. Firstly, the score can only be calculated for an entire QA group, i.e. a query and all the associated memories. To address this, we modify the batching strategy so that each batch contains one query and all the associated memories. Secondly, Eq. \ref{eq:pg} algorithm is hard to optimize especially if the optimization starts from scratch. To resolve this problem, we use curriculum learning~\cite{Bengio2009} under a multi-task learning (MTL) framework. We firstly train the model using Eq. \ref{eq:cs} to kick-start training. We then start training the model using the following  MTL at a reduced learning rate:
% \vspace*{-5pt}
\begin{equation}\label{eq:mix}
  \mathcal{L}(\theta) =   (1- \lambda)  \cdot  \mathcal{L}_{cs}(\theta) + \lambda \cdot  \mathcal{L}_{re}(\theta)
\end{equation}

The new batching strategy whereby a batch consists of an entire QA group is used when the optimization function is set to Eq. \ref{eq:mix}.  $\lambda$ is a hyper-parameter and is determine using random search on validation set.  The reward function is explained in more detail in Sec. \ref{sec:rew_d}.

In addition to aforementioned challenges, Eq. \ref{eq:pg} and in general REINFORCE~\cite{Williams1992} suffers from high variance given its inherent nature of noisy gradient estimates. Selecting the right reward function plays an important role to reduce the variance of gradient estimator \cite{RanzatoCAZ15}. Motivated by this observation and previous works~\cite{RanzatoCAZ15, Rennie2017}, we use the exact score at test time to \textit{baseline} Eq. \ref{eq:pg}:

\begin{equation}\label{eq:base}
  \nabla_{\theta} \mathcal{L}_{\text{re}}(\theta) = (R(a^{M_q}) - \mathtt{F1}(M_q)) \cdot \nabla_{\theta} \text{log }p_{\theta} (M_q | q)
\end{equation}
where $\mathtt{F1}$ is the exact scoring function that is used during test time. We choose an action according to

%\begin{align}\label{eq:maxaction}
%\hat{a} = \{ i \in (m_i, q) | \text{argmax}_{a_j} p(a_j|m_i,q; \theta) \\
%\text{when} ~p(a_j|m_i,q; \theta) > \zeta ~ \text{else} ~ \text{irrelevant} \}
%\end{align}

\begin{equation}
  \label{eq:act_threshold}
  \text{ $\hat{a}_i$ } =
\begin{cases}
    1   & \text{if }   c_i ~\text{is relevant} ~ \&  ~p_{c_i} \geq \zeta   \\
    0              & \text{otherwise}
\end{cases}
\end{equation}
where
\begin{align*}
c_i = \text{argmax}_{a_j} p_{\theta}(a_j|m_i,q)
\end{align*}
is the greedy output of the model, and $\hat{a}$, i.e. $ \hat{a} = \{ \hat{a} \}_{i=1}^{|M_q|} $, is the set of predications for a question and its memories and $\zeta$ is the confidence threshold of the predictions. Using these predictions, $\mathtt{F1}(M_q)$ can be easily calculated for a question and memory group. This method not only helps to reduce the variance by \textit{baselining} Eq. \ref{eq:pg} but also helps the model to make predications with high confidence, given $\zeta$. The ability to directly specify the predication confidence as part of the objective is a distinct advantage over previous methods \cite{PastorPellicer}.

\subsubsection{Reward Function design}\label{sec:rew_d}
Designing an effective reward strategy for use in Eq. \ref{eq:pg} is critical for successful training. This also applies to reinforcement learning in general. Using the vanilla $F1$ score as the reward can have several side-effects. For example, if all the predictions of the model are incorrect, then the $F1$ score becomes zero and, as a result, the loss function become zero and no errors are backpropagted to the network. To address these issues, we define a modified reward function as follows:

\[
   \text{reward} =
\begin{cases}
     1.0   & \text{if }   \text{A}~ \&~ \forall \text{P}  \\
     -0.1   & \text{if }  \text{A}~ \&~ \forall \text{R}  \\
      \textit{accuracy}  & \text{if }   \text{A}~ \&~ \exists \text{P}  \\
    -0.5   & \text{if } \textit{tp} ==0 \\
    -0.01  & \text{if } 0 \leq \mathtt{F1}  \le 0.2  \\
    \mathtt{F1}              & \text{otherwise}
\end{cases}
\]

where $A$ means there are no `relevant'  ground truth labels in the QA group, $P$ means hypothesized `irrelevant' label is correct,  $R$ means hypothesized `irrelevant' label is incorrect, $tp$ denotes number of true positives in QA group, and \textit{accuracy} is the classification accuracy.
\vspace*{-10pt}

%% file: experiments.tex
% !TEX encoding = UTF-8 Unicode

\section{EXPERIMENTS}
\label{sec:experiments}

\subsection{Datasets}
Our data consists of a total approximately 20,000 QA groups divided into the datasets `train', `dev', `TEST-1' and `TEST-2'. Each QA group contains one question. For `TEST-2', the question is an utterance chosen at random from experimental personal assistant logs in which the user has asked the assistant to retrieve a personal memory. For `TEST-1', the user question was typed in by an annotator to resemble an actual user utterance. Table \ref{table:dataset_qa_counts} shows the approximate number of QA groups per dataset in thousands (K). \\

\iffalse
% Table with absolute numbers is commented out to satisfy legal requirements.
\begin{table}[!ht]
\begin{tabular}{ l | c | c }
\bf dataset & \bf  number of QA groups & \bf  number of answers \\ \hline
train & 14,472 & 308,823 \\
dev & 1,236 & 60,883 \\
TEST-1 & 7,659 & 105,365 \\
TEST-2 & 3,031 & 150,645 \\
all & 26,398 & 625,716 \\
\end{tabular}
\caption{number of QA groups and  answers per dataset}\label{table:dataset_qa_counts}
\end{table}
\fi

\begin{table}[!ht]
\begin{tabular}{ l | c | c }
\bf dataset & \bf  number of QA groups & \bf  number of answers \\ \hline
train & $\sim$14K & $\sim$308K \\
dev & $\sim$1K & $\sim$61K \\
TEST-1 & $\sim$8K & $\sim$105K \\
TEST-2 & $\sim$3K & $\sim$151K \\
all & $\sim$26K & $\sim$626K \\
\end{tabular}
\caption{Approximate number of QA groups per dataset}\label{table:dataset_qa_counts}
\end{table}

The answers in each QA group consist of anywhere between 1 to 81 memories which a user had asked the assistant to remember. With the exception of the manually entered questions in `TEST-1', the questions  and memories are the output of the assistant's speech recognition engine and, as a result,  contain speech recognition errors. Each of the memories has been manually annotated as being relevant or irrelevant to the question in its QA group.

%An example of a QA group can be found below. \textcolor{Green}{$\checkmark$} means 'relevant' and \textcolor{Red}{$\times$} 'irrelevant' to the question.\\

%\fbox{\begin{minipage}{23em}
%\textbf{Question}:
%what did i do with ben's cell phone  \\

%\textbf{Answers}:
%\begin{enumerate}[itemsep=-0.5mm]
%  \item \textcolor{Green}{$\checkmark$} i gave benny's cell in for repairs at the store on first street
%  \item \textcolor{Green}{$\checkmark$} i left ben's iphone on the kitchen table
%  \item \textcolor{Green}{$\checkmark$} i sent bennie's old phone to mat
%  \item \textcolor{Red}{$\times$} ben wants a new cell phone for his birthday
%  \item \textcolor{Red}{$\times$} dad's cell is an iphone eight
%  \item \textcolor{Red}{$\times$} the screen of benjamin's phone is broken
%\end{enumerate}
%\end{minipage}} \\\\

 As is apparent in Table \ref{table:memory_counts}, the number of memories per QA group differs significantly across datasets. This is because the number of memories each user has stored varies greatly and the maximum number of memories annotated per QA group varied between annotation groups.\\

\begin{table}[!ht]
\centering
\begin{tabular}{ | l | c | c | c | c | } \hline
\bf dataset  & \bf min & \bf max & \bf mean & \bf std. dev.  \\  \hline
train & 1.0  & 80.0 & 21.34 & 19.35 \\
dev &18.0 & 81.0 & 49.26 & 29.91 \\
TEST-1 & 1.0 & 30.0 & 13.76 & 10.45 \\
TEST-2 & 18.0 & 81.0 & 49.70 & 29.91\\\hline
%all & 1.0 & 81.0 & 23.70 & 22.83\\ \hline
\end{tabular}
\caption{Minimum, maximum and mean number of memories across QA groups in each dataset.}\label{table:memory_counts}
\end{table}

There is a significant class imbalance between relevant and irrelevant memories across datasets as most of a user's memories are not relevant to a given question. Table  \ref{table:relevant_label_percentages} shows the percent of memories with a `relevant'-label in each QA group averaged over all the QA groups in dataset. Because of this imbalance, relevant answers were upsampled using weighted sampling during training to create batches with roughly the same number of relevant and irrelevant examples.\\

\begin{table}[!ht]
\centering
\begin{tabular}{ | c | c | }  \hline
\bf dataset & \% \bf relevant memories \\ \hline
train  & 15.08\% \\  \hline
dev & 4.54\% \\  \hline
TEST-1 & 20.94\% \\  \hline
TEST-2  &  4.59\% \\ \hline
% all & 15.08\% \\ \hline
\end{tabular}
\caption{Percentage of memories with `relevant'-label in each QA group averaged over all QA groups in dataset}\label{table:relevant_label_percentages}
\end{table}

Each question and answer also undergoes a preprocessing step in which contractions and abbreviations are expanded, e.g. ``doesn't'' $\rightarrow$ ``does not'', ``wanna'' $\rightarrow$ ``want to'', common question carrier phrases are removed, e.g. ``can you remember what'', ``please tell me who'' and stop words with little meaning are deleted, e.g. ``did'', ``does", ``is''. Such preprocessing makes the wording more consistent across utterances and removes words and phrases with little or no semantic content. It decreases the average number of tokens in questions from 6.6 to 3.8 and in answers from 4.2 to 3.7. The change in number of tokens for each data is listed in Table \ref{table:num_tokens_per_dataset}.  \\

\begin{table}[!ht]
\centering
\begin{tabular}{ | l | c | c | c | c |}  \hline
  & \multicolumn{2}{c |}{\bf raw} & \multicolumn{2}{c |}{\bf preprocessed} \\  \hline
 \bf dataset & \bf question & \bf answer & \bf question & \bf answer \\  \hline
train & 6.8 & 3.7 & 4.0 & 3.3 \\
dev & 7.0 & 3.8 & 4.1 & 3.4 \\
TEST-1 & 6.0 & 6.7 & 3.3 & 5.6 \\
TEST-2 & 7.1 & 3.8 & 4.2 & 3.4 \\ \hline
%all & 6.6 & 4.2 & 3.8 & 3.7 \\  \hline
\end{tabular}
\caption{Average number of tokens per question and per answer in each dataset}\label{table:num_tokens_per_dataset}
\end{table}
\vspace{-25pt}

\subsection{Evaluation metrics}
Each model  was evaluated on the test sets `TEST-1', where the questions were typed in by annotators, and `TEST-2', where questions are the output of an ASR engine. For each QA group in the dataset, the precision, recall and $F1$ score were calculated by comparing the relevance labels assigned by annotators with the hypotheses returned by the model. The precision, recall and $F1$ score of all QA groups in each dataset were then averaged to give the average precision, average recall and average $F1$ score for the entire dataset. Ranking of memories was not considered.

\subsection{Model specifications}

%We report results on the $\mathtt{TEFF}$,  $\mathtt{TEFFCH}$, $\mathtt{SARD}$ and $\mathtt{SARDCH}$ models.
We report results on the $\mathtt{TEFF}$ and  $\mathtt{TEFFCH}$. We used the Adam optimizer \cite{kinga2015method} to train all models with Eq. \ref{eq:cs} and use a constant learning rate of $0.001$. When the training objective is switched to Eq. \ref{eq:mix}, we adopt a different batching scheme and decay the learning rate by a factor of $0.1$. We use the $\mathtt{ReLu}$ activation function throughout our models. We use L$2$ weight decay to train our models and apply dropout at a rate of $0.1$ across all models. We use a batch size of $128$ and pick the model with the best performance on the validation set. We use random search for hyperamater tuning to determine the best model configuration. We use two variations of batching for our experiments. For models trained using Eq. \ref{eq:cs}, we construct a batch of query-memory pairs using random sampling, but oversample from the positive examples to ensure a $1:1$ ratio of positive and negative examples in every batch.
However, when using the MTL objective in Eq. \ref{eq:mix}, we batch all memories for a given question to compute the $F1$ score for the given QA group. For models that only consume word embeddings, we batch the queries and memories to a maximum sentence length, and all sentences smaller than this length were padded using a <$\mathtt{PAD}$> token. For all our models, we use $300$ dimensional pre-trained $\mathtt{fastText}$ word vectors \cite{mikolov2018advances}, trained on Wikipedia data from $2017$, news datasets from \texttt{statmt.org} from $2007$-$2016$ as well as the UMBC corpus \cite{han2013umbc_ebiquity}. We found that a maximum utterance length of $10$ was sufficient to ensure good results, given that the average query length was much shorter (after preprocessing). For models with a $\mathtt{CharCNN}$ module, the character level inputs were also padded at the word-level using a maximum word length of $8$.

Our best $\mathtt{TEFF}$ model is 2-layer network with $694$ units each. The best $\mathtt{TEFFCH}$ model employs two convolution layers of with a kernel size of $1$ and $2$, and $128$ filters each, followed by a linear layer outputting $108$ dimensional $\mathtt{CharCNN}$ embeddings, concatenated together with pre-trained word vectors to give $408$ dimensional word representations. The concatenated embeddings are then processed through a 2-layer network with $736$ units each.  All models have a final softamx layer to output a distribution over two classes.

%The $\mathtt{CharCNN}$ module in $\mathtt{SARDCH}$ model is comprised of a single convolutional layer of width $1$, with $128$ features, followed by a linear layer outputting $160$ dimensional $\mathtt{CharCNN}$ embeddings.
%For the $\mathtt{SARD}$ and $\mathtt{SARDCH}$ models, we use two highway layers of size $300$ and $460$ each, $4$ heads of $75$ and $115$ dimensional attentional states, $936$ and $354$ units for the position-wise feed-forward network, followed by two linear layers of $300$ and $460$ units each, respectively.

\subsection{Smooth Approximation}
\label{subsec:smooth}
%\footnote{After conducting experiments for smooth $F1$ approximation, we found the independent work in \cite{PastorPellicer} to be similar in principal to our %approach}

In order to have a comprehensive evaluation of our proposed method, we compare our method with \cite{PastorPellicer}, wherein a smooth approximation of the $F1$ score was proposed. This smooth objective is differentiable and is formulated as follows:

\begin{align}\label{eq:f1}
\mathcal{L}_{fs}(\theta) &= -\sum\limits_{j } \mathcal{F}_{\theta}(M_{q_j} |q_j) \\
\text{PR}_{\theta} &= \frac{  \sum_i{   \text{log }p_{\theta}  \mathbb{I}_{tp}(q_j,m_i) }    } {\sum_i{   \text{log }p_{\theta}  \mathbb{I}_{tp }(q_j,m_i) }  + \sum_i{   \text{log }p_{\theta}  \mathbb{I}_{fp}(q_j,m_i) }  } \\
\text{RE}_{\theta} &= \frac{  \sum_i{   \text{log }p_{\theta}  \mathbb{I}_{tp}(q_j,m_i) }   } {\sum_i{  \text {log }p_{\theta}  \mathbb{I}_{tp }(q_j,m_i) }  + \sum_i{   \text{log }p_{\theta}  \mathbb{I}_{fn}(q_j,m_i) }  }   \\
 \mathcal{F}_{\theta} &= 2  \cdot \frac{ \text{PR}_{\theta}  \cdot \text{RE}_{\theta} }{\text{PR}_{\theta}  + \text{RE}_{\theta} }
\end{align}
where $\mathbb{I}_{tp}$ is the indicator function for true positives, $\mathbb{I}_{fp}$ for false positives, and $\mathbb{I}_{fn}$ for false negatives. Since $\mathcal{F}$ is a differentiable function and is parameterized by $\theta$, we can directly optimize for it during training. We followed the same steps as for the MTL loss (Eq. \ref{eq:mix}), i.e. the different batching strategy and decayed learning rate.

One of the main advantages of our proposed method compared to the smooth formulation, is that we can directly enforce the confidence level in the prediction as in Eq. \ref{eq:act_threshold}. Moreover, the results show that our proposed method significantly outperforms the above smooth formulation.

%\subsection{Noisy Training Data}

%Regularization is a vital component when training neural networks, as models have high capacity and tend to overfit the training data. We investigate the effect of using a larger, noisy training set, on model generalization, by using a simple heuristic to bootstrap new data collection from an existing system. Generating training data in this fashion can prove to be an effective augmentation scheme which can increase model performance and mitigate the problem of class imbalance, by increasing the diversity of samples present. However, this technique can also backfire and lead to instabilities during training by corrupting the data with too much noise, depending on the precision of the model used to generate labels. We use the results of a high-precision lexical search engine to determine if a memory is considered relevant to the query. Using this automated approach, we were able to annotate $\sim 125$K QA groups. We then use this unsupervised data to augment our training set and report results on models trained on this bigger dataset in Figure ``<FILL IN LATER>''.

% We also experimented with training word vectors on our augmented corpus, and using these vectors together with pre-trained embeddings, as a means of encoding task-specific information into our inputs. We found that this did not yield favorable results and hence do not report results for the same.

\section{Results and Analysis}
\label{sec:analysis}

\begin{figure}[tb!]
  \centering
  \begin{subfigure}[b]{0.49\columnwidth}
    % \centering
    \includegraphics[width=\linewidth]{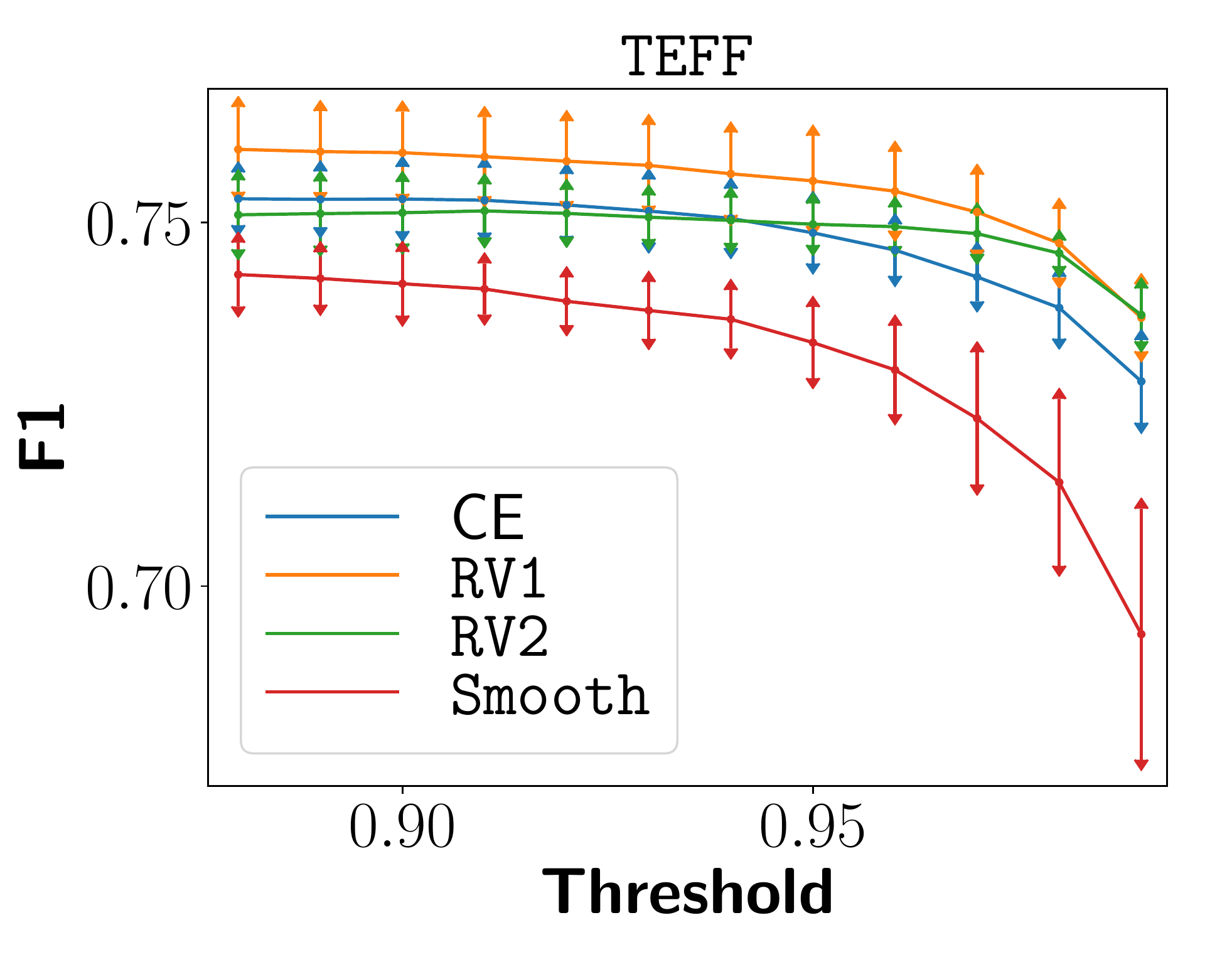}
    % \caption{$\mathtt{TEFF}$}
  \end{subfigure}
  \begin{subfigure}[b]{0.49\columnwidth}
    % \centering
    \includegraphics[width=\linewidth]{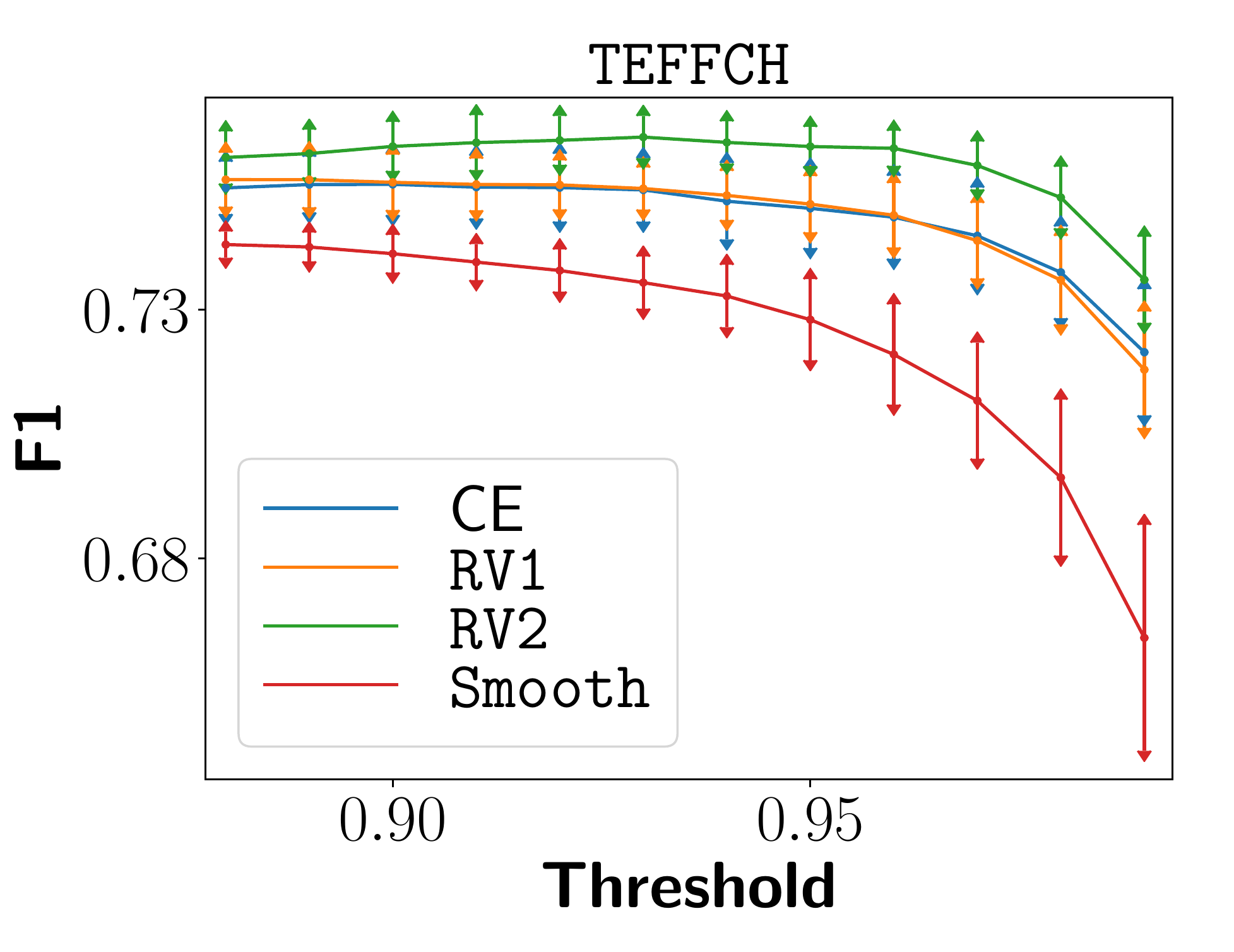}
    % \caption{$\mathtt{TEFFCH}$}
  \end{subfigure}
  \caption{Plot comparing `TEST-1' performance for the various aforementioned training objectives. The error bars were generated by running multiple trials of the same model, with different seeds. }
  \label{fig:relative}
\end{figure}

\begin{table}[!t]
  \begin{center}
    \resizebox*{\columnwidth}{!}{
    \begin{tabular}{l|c|ccc|c}
      \hline
      \bf Model & \bf Objective & $t=0.97$ & $t=0.98$ & $t=0.99$ & \bf \# Params \\
      \hline
      \multirow{3}{*}{$\mathtt{TEFF}$} & $\mathtt{RV1}$ & 2.283\% & 2.295\% & 2.259\% & \multirow{3}{*}{0.70M}\\
      & $\mathtt{RV2}$ & 1.075\% & 1.772\% & 2.766\% &\\
      & $\mathtt{Smooth}$ & -1.673\% & -1.832\% & -2.405\% &\\
      \hline
      \multirow{3}{*}{$\mathtt{TEFFCH}$} & $\mathtt{RV1}$ & 1.807 \% & 2.896 \% & 5.016\% & \multirow{3}{*}{0.89M}\\
      & $\mathtt{RV2}$ & 3.714\% & 4.732\% & 5.239\% &\\
      & $\mathtt{Smooth}$ & -1.557\% & -1.413\% & -1.407\% &\\
      \hline
    \end{tabular}
    }
  \end{center}
  \caption{\label{tab:relative} Table showing the relative improvement in $F1$ score achieved using various training objectives across different thresholds, compared to the cross-entropy baseline, on `TEST-1'. The model size is given by \textbf{\# Params}.}
\end{table}

A key aspect of our proposed objective is that the model is made aware of the type of errors\footnote{Type I - False positive or Type II - False negative} in its prediction, and can use this information to trade-off the number of false positives and false negatives, for an optimal $F1$ score. In our experiments, when training with Eq. \ref{eq:cs}, the model is able predict, with reasonable accuracy, the relevant memories, for various query types. However, for some QA groups that contain noise in the form of ASR errors or otherwise complex queries, this accuracy is lower as expected. For these challenging cases, our method encourages the network to balance the number of false positives and false negatives, to avoid predictions with high precision and low recall (and vice-versa), but rather maintain an optimal balance of the two.

\begin{figure}[htb!]
  \centering
  \includegraphics[width=\columnwidth]{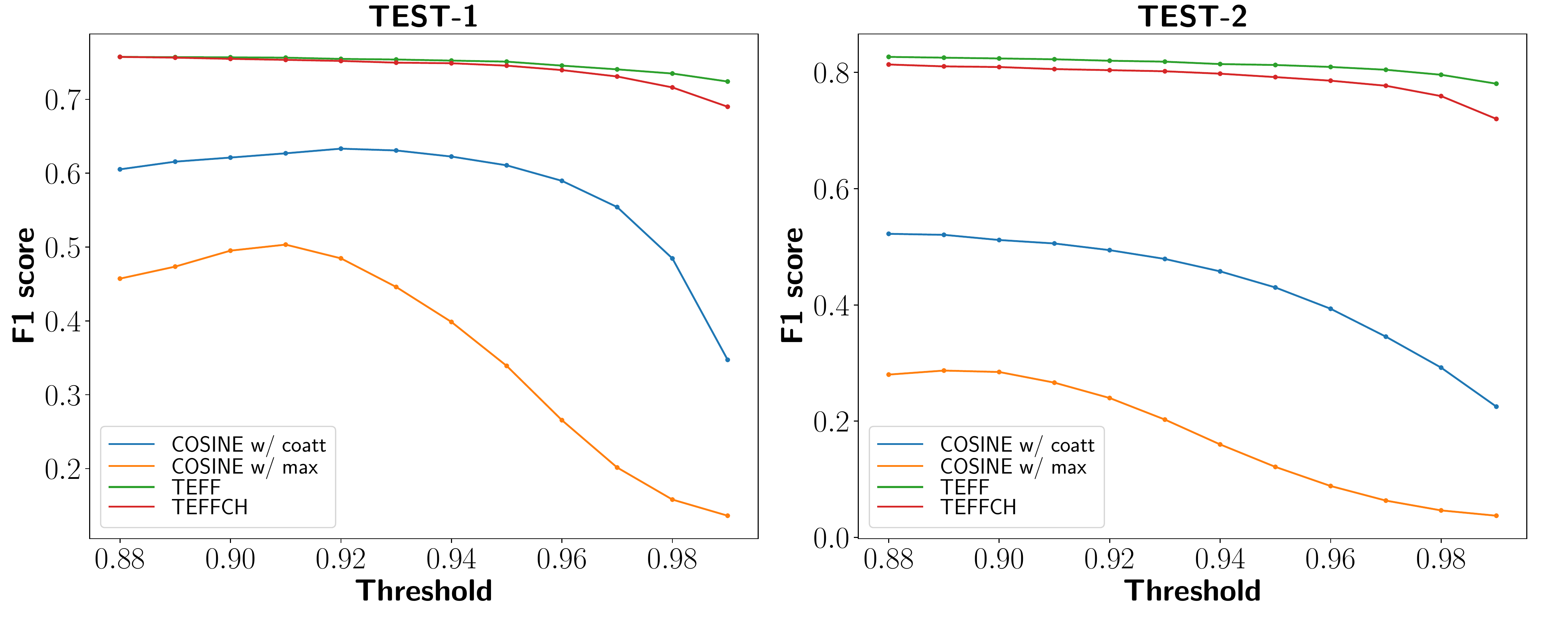}
  \caption{Performance of various models across our test sets. The $\mathtt{COSINE}$ models is a parameter-free \acrfull{BOW} model that consumes word embeddings, which are aggregated using  1) $\mathtt{COATT}$, which uses a coattention encoding similar to the one presented in \cite{Xiong2016DynamicCN} and 2) $\mathtt{MAX}$, which simply does a max-pooling across time to produce fixed dimensional vectors. These words are computed for both the query and memories, and a memory is deemed relevant if the cosine similarity between the query and memory vectors exceeds a certain threshold. }
\label{fig:baselines}
\end{figure}

\begin{figure}[htb!]

\includegraphics[width=\linewidth,]{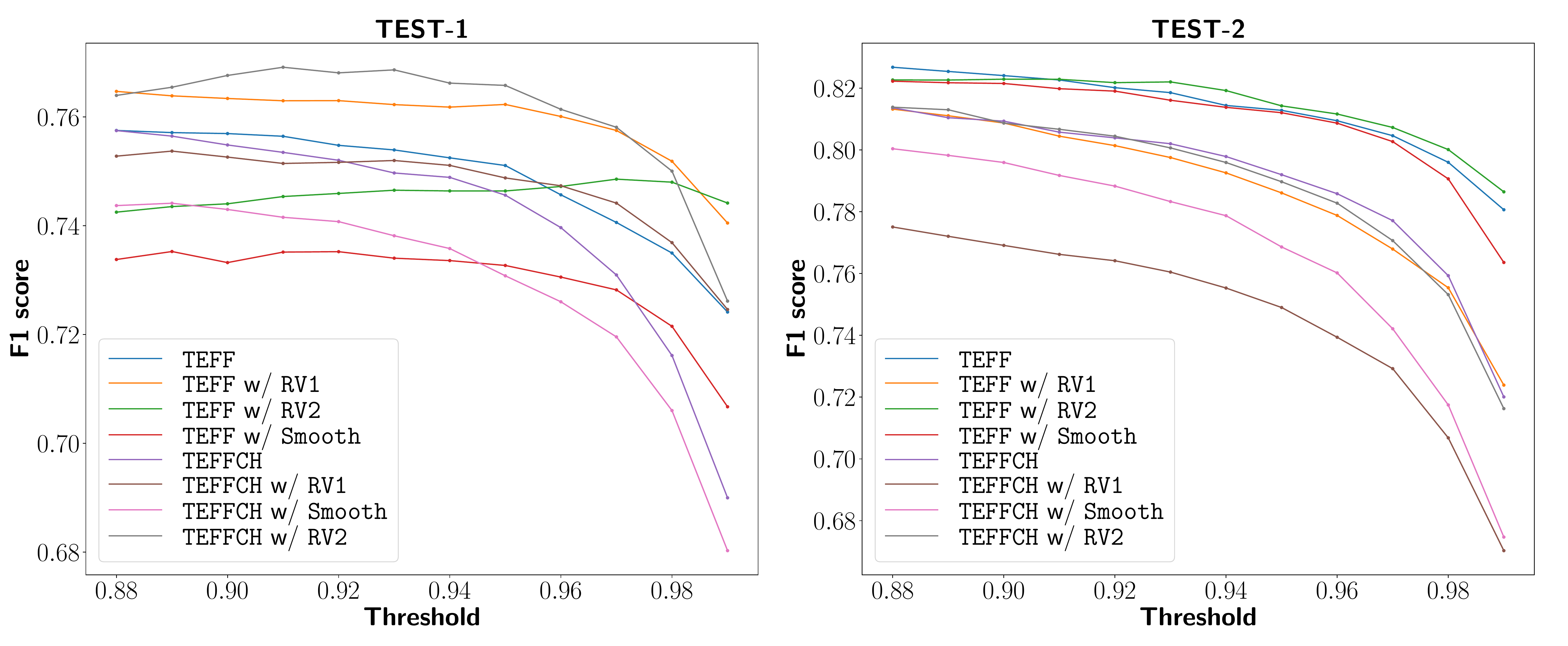}
\caption{Plot showing the evolution of $F1$ scores across various thresholds, and across models $\mathtt{TEFF}$ and $\mathtt{TEFFCH}$. Both models show an improvement when optimizing using Eq. \ref{eq:mix}, while the $\mathtt{Smooth}$ $F1$ objective as described in section \ref{subsec:smooth} performs worse than our baseline model (Eq. \ref{eq:cs}). $\mathtt{RV1}$ refers to models trained using Eq. \ref{eq:gpg}, whereas $\mathtt{RV2}$ refers to the reduced variance \textit{baseline} described in Eq. \ref{eq:base}. }
\label{fig:F1}
\end{figure}

We investigate the gains realized when optimizing using our MTL objective in Eq. \ref{eq:mix} and compare it to using the standard loss formulation (Eq. \ref{eq:cs}) and smooth loss (Sec. \ref{subsec:smooth}) as a training objective. As shown in Table \ref{tab:relative} and Figure \ref{fig:relative}, we find that our proposed optimization objective, infact leads to an improvement in $F1$ score for this QA task. In Figure \ref{fig:relative}, the error bars were generated by setting different seeds for each trial. However, we fix the curriculum learning strategy beforehand, and do not tune it for each trial. This may be sub-optimal for the performance of the model. Tuning this curriculum learning strategy per trial will likely show futher gains.
Figures \ref{fig:baselines} and \ref{fig:F1} show the performance of various models trained with different objectives, across our test sets. In our results, $\mathtt{RV1}$ and $\mathtt{RV2}$ refer to training using Eqs. \ref{eq:gpg} and \ref{eq:base} respectively, and $\mathtt{Smooth}$ refers to the objective described in Section \ref{subsec:smooth}.

We observe that the underlying structure of `TEST-1' and `TEST-2' are different, and hence show different gains in performance. Both test sets and the training set have on average, about $3$ relevant memories per QA group, but `TEST-2' has more total memories per QA group, and shorter memories, on average (Table \ref{table:num_tokens_per_dataset}). We find that `TEST-2' shows smaller gains relative to `TEST-1', when using our objective, as `TEST-2' contains shorter utterances, which provide less context to distinguish different memories. However, on `TEST-1', our objective was able to learn a stronger semantic model which is able to adapt to correctly identify paraphrasing, ASR errors and long-range context in complex utterances (Figure \ref{fig:F1}).

% For a query with a large number of memories, the reward is therefore scaled to small in magnitude compared to a query with a lot fewer memories. This is possibly an explanation as to why we observe larger gains in performance on `TEST-1', which is similar in distribution of the number of memories to the training data, relative to `TEST-2', which does not benefit as much. One possible strategy would be to normalize the reward by a function of the number of relevant memories, as opposed to the total number of memories. This technique would make the optimization update independent of the total size of the QA group and encourage the network to allocate its capacity to adaptively optimize for precision and recall, given the ratio of positive and negative labels present.

Moreover, we expected larger gains using $\mathtt{CharCNN}$, but this was not always the case. We hypothesize that this is due to the short nature of utterances in our \textit{training} data, which are only $4$ tokens long, on average (see Table \ref{table:num_tokens_per_dataset}), and can be suitably encoded using pre-trained word vectors. The pre-trained word vectors are trained on a much larger corpus and generalize well as they aggregate contextual information from multiple domains. On the other hand, our $\mathtt{CharCNN}$ module is not processed through a sequential network, and hence lacks inter-word context. This module is also trained on our task-specific loss which may not be optimal for learning such embeddings, and can depict quite pathological behviour, given the size of the dataset and the length of utterances. This can cause the model to produce overloaded representations and make it prone to overfitting, thereby harder to train. We hypothesize that with less pre-processing, the additional context available could mitigate some of the issues exhibited by the $\mathtt{CharCNN}$ module, leading to futher gains.

%% file: conclusion.tex
\section{Conclusion}

In this paper, we present an end-to-end system for spoken personal question answering. Moreover, we propose a novel objective function to directly optimize the $F1$ measure for our information retrieval task. By directly optimizing the $F1$-score, we can take into account the predicted labels of \textit{all} answers simultaneously. It also enables us to take the types of errors into consideration during optimization, e.g. number of false positives, number of false negatives. Furthermore, our proposed objective can mitigate the effects of class imbalance and noisy data in the form of ASR errors. Our extensive experimentation shows that the aforementioned approaches deliver benefits to system performance. We also analyze the impact of our methods on different datasets with varying structure.

%% file: main.bbl
\begin{thebibliography}{10}

\bibitem{hinton2012deep}
Geoffrey Hinton, Li~Deng, Dong Yu, George~E Dahl, Abdel-rahman Mohamed, Navdeep
  Jaitly, Andrew Senior, Vincent Vanhoucke, Patrick Nguyen, Tara~N Sainath,
  et~al.,
\newblock ``Deep neural networks for acoustic modeling in speech recognition:
  The shared views of four research groups,''
\newblock {\em IEEE Signal Processing Magazine}, vol. 29, no. 6, pp. 82--97,
  2012.

\bibitem{DahlAcero}
G.~E. Dahl, D.~Yu, L.~Deng, and A.~Acero,
\newblock ``Context-dependent pre-trained deep neural networks for
  large-vocabulary speech recognition,''
\newblock {\em IEEE Transactions on Audio, Speech, and Language Processing},
  vol. 20, no. 1, pp. 30--42, Jan 2012.

\bibitem{rf2017NIPSW}
Rasool Fakoor, Xiaodong He, Ivan Tashev, and Shuayb Zarar,
\newblock ``Reinforcement learning to adapt speech enhancement to instantaneous
  input signal quality,''
\newblock in {\em NIPS 2017, Machine Learning for Audio Signal Processing
  workshop}, 2017.

\bibitem{rfICASSP2018}
Rasool Fakoor, Xiaodong He, Ivan Tashev, and Shuayb Zarar,
\newblock ``Constrained convolutional-recurrent networks to improve speech
  quality with low impact on recognition accuracy,''
\newblock {\em CoRR}, vol. abs/1802.05874, 2018.

\bibitem{zettlemoyer2012learning}
Luke~S Zettlemoyer and Michael Collins,
\newblock ``Learning to map sentences to logical form: Structured
  classification with probabilistic categorial grammars,''
\newblock {\em arXiv preprint arXiv:1207.1420}, 2012.

\bibitem{Mori2008}
R.~De Mori, F.~Bechet, D.~Hakkani-Tur, M.~McTear, G.~Riccardi, and G.~Tur,
\newblock ``Spoken language understanding,''
\newblock {\em IEEE Signal Processing Magazine}, vol. 25, no. 3, pp. 50--58,
  May 2008.

\bibitem{weston2015towards}
Jason Weston, Antoine Bordes, Sumit Chopra, Alexander~M Rush, Bart van
  Merri{\"e}nboer, Armand Joulin, and Tomas Mikolov,
\newblock ``Towards ai-complete question answering: A set of prerequisite toy
  tasks,''
\newblock {\em arXiv preprint arXiv:1502.05698}, 2015.

\bibitem{Xiong2016DynamicCN}
Caiming Xiong, Victor Zhong, and Richard Socher,
\newblock ``Dynamic coattention networks for question answering,''
\newblock {\em ICLR}, 2017.

\bibitem{Sordoni2016IterativeAN}
Alessandro Sordoni, Phillip Bachman, and Yoshua Bengio,
\newblock ``Iterative alternating neural attention for machine reading,''
\newblock {\em CoRR}, vol. abs/1606.02245, 2016.

\bibitem{young2010hidden}
Steve Young, Milica Ga{\v{s}}i{\'c}, Simon Keizer, Fran{\c{c}}ois Mairesse,
  Jost Schatzmann, Blaise Thomson, and Kai Yu,
\newblock ``The hidden information state model: A practical framework for
  pomdp-based spoken dialogue management,''
\newblock {\em Computer Speech \& Language}, vol. 24, no. 2, pp. 150--174,
  2010.

\bibitem{LiMRJGG16}
Jiwei Li, Will Monroe, Alan Ritter, Dan Jurafsky, Michel Galley, and Jianfeng
  Gao,
\newblock ``Deep reinforcement learning for dialogue generation.,''
\newblock in {\em EMNLP}, 2016, pp. 1192--1202.

\bibitem{Sarikaya}
Ruhi Sarikaya,
\newblock ``The technology behind personal digital assistants: An overview of
  the system architecture and key components,''
\newblock {\em IEEE Signal Processing Magazine}, vol. 34, no. 1, pp. 67--81,
  Jan 2017.

\bibitem{rajpurkar2016squad}
Pranav Rajpurkar, Jian Zhang, Konstantin Lopyrev, and Percy Liang,
\newblock ``Squad: 100,000+ questions for machine comprehension of text,''
\newblock {\em arXiv preprint arXiv:1606.05250}, 2016.

\bibitem{mitra2017neural}
Bhaskar Mitra and Nick Craswell,
\newblock ``Neural models for information retrieval,''
\newblock {\em arXiv preprint arXiv:1705.01509}, 2017.

\bibitem{wang2017bilateral}
Zhiguo Wang, Wael Hamza, and Radu Florian,
\newblock ``Bilateral multi-perspective matching for natural language
  sentences,''
\newblock {\em arXiv preprint arXiv:1702.03814}, 2017.

\bibitem{seo2016bidirectional}
Minjoon Seo, Aniruddha Kembhavi, Ali Farhadi, and Hannaneh Hajishirzi,
\newblock ``Bidirectional attention flow for machine comprehension,''
\newblock {\em arXiv preprint arXiv:1611.01603}, 2016.

\bibitem{QuocOptim2007}
Quoc~V. Le and Alexander~J. Smola,
\newblock ``Direct optimization of ranking measures,''
\newblock {\em CoRR}, vol. abs/0704.3359, 2007.

\bibitem{PastorPellicer}
Joan Pastor-Pellicer, Francisco Zamora-Mart\'{\i}nez, Salvador Espa\~{n}a
  Boquera, and Mar\'{\i}a~Jos{\'e} Castro-Bleda,
\newblock ``F-measure as the error function to train neural networks,''
\newblock in {\em Proceedings of the 12th International Conference on
  Artificial Neural Networks: Advances in Computational Intelligence - Volume
  Part I}. 2013, IWANN'13, pp. 376--384, Springer-Verlag.

\bibitem{Xu2016ExpectedFT}
Wenduan Xu, Michael Auli, and Stephen Clark,
\newblock ``Expected f-measure training for shift-reduce parsing with recurrent
  neural networks,''
\newblock in {\em HLT-NAACL}, 2016.

\bibitem{Williams1992}
Ronald~J. Williams,
\newblock ``Simple statistical gradient-following algorithms for connectionist
  reinforcement learning,''
\newblock {\em Mach. Learn.}, vol. 8, no. 3-4, pp. 229--256, May 1992.

\bibitem{RanzatoCAZ15}
Marc'Aurelio Ranzato, Sumit Chopra, Michael Auli, and Wojciech Zaremba,
\newblock ``Sequence level training with recurrent neural networks,''
\newblock {\em ICLR}, vol. abs/1511.06732, 2016.

\bibitem{Rennie2017}
Steven~J. Rennie, Etienne Marcheret, Youssef Mroueh, Jarret Ross, and Vaibhava
  Goel,
\newblock ``Self-critical sequence training for image captioning,''
\newblock in {\em CVPR}, pp. 1179--1195. 2017.

\bibitem{kim2016character}
Yoon Kim, Yacine Jernite, David Sontag, and Alexander~M Rush,
\newblock ``Character-aware neural language models.,''
\newblock in {\em AAAI}, 2016, pp. 2741--2749.

\bibitem{jozefowicz2016exploring}
Rafal Jozefowicz, Oriol Vinyals, Mike Schuster, Noam Shazeer, and Yonghui Wu,
\newblock ``Exploring the limits of language modeling,''
\newblock {\em arXiv preprint arXiv:1602.02410}, 2016.

\bibitem{ZarembaS15}
Wojciech Zaremba and Ilya Sutskever,
\newblock ``Reinforcement learning neural turing machines,''
\newblock {\em CoRR}, vol. abs/1505.00521, 2015.

\bibitem{Bengio2009}
Yoshua Bengio, J{\'e}r\^{o}me Louradour, Ronan Collobert, and Jason Weston,
\newblock ``Curriculum learning,''
\newblock in {\em ICML}, New York, NY, USA, 2009, ICML '09, pp. 41--48, ACM.

\bibitem{kinga2015method}
D~Kinga and J~Ba Adam,
\newblock ``A method for stochastic optimization,''
\newblock in {\em International Conference on Learning Representations (ICLR)},
  2015, vol.~5.

\bibitem{mikolov2018advances}
Tomas Mikolov, Edouard Grave, Piotr Bojanowski, Christian Puhrsch, and Armand
  Joulin,
\newblock ``Advances in pre-training distributed word representations,''
\newblock in {\em LREC}, 2018.

\bibitem{han2013umbc_ebiquity}
Lushan Han, Abhay~L Kashyap, Tim Finin, James Mayfield, and Jonathan Weese,
\newblock ``Umbc\_ebiquity-core: semantic textual similarity systems,''
\newblock in {\em Second Joint Conference on Lexical and Computational
  Semantics (* SEM), Volume 1: Proceedings of the Main Conference and the
  Shared Task: Semantic Textual Similarity}, 2013, vol.~1, pp. 44--52.

\bibitem{Transfomer2017}
Ashish Vaswani, Noam Shazeer, Niki Parmar, Jakob Uszkoreit, Llion Jones,
  Aidan~N Gomez, \L~ukasz Kaiser, and Illia Polosukhin,
\newblock ``Attention is all you need,''
\newblock in {\em NIPS}, pp. 5998--6008. 2017.

\bibitem{SrivastavaGS15}
Rupesh~Kumar Srivastava, Klaus Greff, and J{\"{u}}rgen Schmidhuber,
\newblock ``Highway networks,''
\newblock {\em CoRR}, vol. abs/1505.00387, 2015.

\bibitem{Highway}
Rupesh~K Srivastava, Klaus Greff, and J\"{u}rgen Schmidhuber,
\newblock ``Training very deep networks,''
\newblock in {\em NIPS}, pp. 2377--2385. 2015.

\bibitem{Qin2010}
Tao Qin, Tie-Yan Liu, and Hang Li,
\newblock ``A general approximation framework for direct optimization of
  information retrieval measures,''
\newblock {\em Information Retrieval}, vol. 13, no. 4, pp. 375--397, Aug 2010.

\bibitem{Krzysztof2011}
Krzysztof~J. Dembczynski, Willem Waegeman, Weiwei Cheng, and Eyke
  H\"{u}llermeier,
\newblock ``An exact algorithm for f-measure maximization,''
\newblock in {\em NIPS}, pp. 1404--1412. 2011.

\bibitem{Dumais2003}
Susan Dumais, Edward Cutrell, JJ~Cadiz, Gavin Jancke, Raman Sarin, and
  Daniel~C. Robbins,
\newblock ``Stuff i've seen: A system for personal information retrieval and
  re-use,''
\newblock in {\em SIGIR}. 2003, pp. 72--79, ACM.

\bibitem{Burgeslambdamart}
Chris~J.C. Burges,
\newblock ``From ranknet to lambdarank to lambdamart: An overview,''
\newblock Tech. {R}ep., June 2010.

\bibitem{berant2013semantic}
Jonathan Berant, Andrew Chou, Roy Frostig, and Percy Liang,
\newblock ``Semantic parsing on freebase from question-answer pairs,''
\newblock in {\em EMNLP}, 2013, pp. 1533--1544.

\bibitem{vaswani2017attention}
Ashish Vaswani, Noam Shazeer, Niki Parmar, Jakob Uszkoreit, Llion Jones,
  Aidan~N Gomez, {\L}ukasz Kaiser, and Illia Polosukhin,
\newblock ``Attention is all you need,''
\newblock in {\em Advances in Neural Information Processing Systems}, 2017, pp.
  5998--6008.

\bibitem{DanielGuo}
Daniel~(Zhaohan) Guo, Gokhan Tur, Scott Wen-tau Yih, and Geoffrey Zweig,
\newblock ``Joint semantic utterance classification and slot filling with
  recursive neural networks,''
\newblock in {\em IEEE SLT}, December 2014.

\end{thebibliography}
